\documentclass[twocolumn]{article}
\usepackage[dvips]{graphicx}
\usepackage{bm}
\usepackage{amsmath}

\begin{document}

\title{Flat-band excitonic states in Kagome lattice \\ on semiconductor surfaces}
\author{$\text{H. Ishii}^{(1)}$, $\text{T. Nakayama}^{(1)}$, $\text{J. Inoue}^{(2)}$ 
\\ (1)Department of Physics, Chiba University, 1-33 Yayoi, Inage, Chiba 263-8522, Japan 
\\ (2)Center for Frontier Science, Chiba University, 1-33 Yayoi, Inage, Chiba 263-8522, Japan}

\maketitle

\section{Introduction}
Recent surface techniques to fabricate and arrange quantum wires on 
the semiconductor surface have enabled to synthesize the artificial 
two-dimensional lattice systems such as square and triangle lattices. 
For example, C. Albrecht $et \ \ al$. produced a square lattice by arranging 
InAs wires on GaInAs (001) surface and showed that a butterfly energy 
spectra is seen in the case of applying the magnetic field. 
[1] Among various lattices, a Kagome lattice is located at the 
special position because of the appearance of complete flat bands 
in the electronic spectra. [2,3] Since the density of states of such 
flat bands is macroscopically large and the electronic correlation 
remarkably works when the half of a flat band is occupied by electrons 
through the application of the gate voltage, the surface ferromagnetism 
is predicted for nonmagnetic semiconductor surfaces [2,3] 
and the experimental challenges to realize a Kagome lattice 
is now in progress. In this view, one can also expect that a 
Kagome lattice shows an exotic optical properties, 
which have never been studied yet. In this paper, 
we investigated the excitonic properties of the Kagome lattice. 
Most remarkable finding is that the exciton binding energy 
on the Kagome lattice is extremely large; larger than that 
in one-dimensional system. By changing the dimensionality of the lattice system, 
it is shown that such large binding energy originates from the flatness of 
the lowest-conduction and highest-valence bands in the Kagome lattice. 
Moreover, by calculating the binding energies of charged exciton and biexciton, 
we show that the excitons are easy to form charged exciton. 
We expect that the present results will give useful 
information for the optical detection of the flat bands.

\section{Model and Method}
 The local spin density functional calculation showed that when quantum wires 
are arranged on the semiconductor surfaces in the way to form a Kagome lattice, 
the lower conduction-band electrons are mainly localized around 
the cross points of the quantum wires. [3] This enables us to reasonably assume 
that the electronic structures of electrons and holes in the lower conduction 
and higher valence bands are well described by employing the tight-binding model, 
where electrons and holes are located on the cross points of quantum wires and 
transfer along the wires. This situation is schematically shown as solid lines 
in Fig.1(b). Then the model Hamiltonian becomes

\[ \resizebox{0.95\hsize}{!}{$\displaystyle H=-\sum_{m,n} t^a_{mn} \hat{a}^{\dagger}_m 
\hat{a}_n -\sum_{m,n} t^b_{mn} \hat{b}^{\dagger}_m \hat{b}_n 
-\sum_{m,n} U^{ab}_{mn} \hat{a}^{\dagger}_m \hat{a}_m \hat{b}^{\dagger}_n \hat{b}_n \ \ ,$} \]
where $\hat{a}_{n}$ and $\hat{b}_{n}$ represent the annihilation operators of 
an electron and a hole at the $n$ site, respectively, and $t^{a}_{mn}$ and 
$t^{b}_{mn}$ are transfer energies of these carriers from the $n$ site to the $m$ site.
$U^{ab}_{mn}$ is the Coulomb attraction energy between an electron and a hole, for which 
we adopted the following form;

\begin{equation*}
 U^{ab}_{mn}=\begin{cases}
               U_0 & r = 0 \\
               U_1/r & r \neq 0 \ \ \ ,
              \end{cases}
\end{equation*} 
where $r_{mn}$is the distance between $m$ and $n$ sites. [4] Here we assumed that the 
lattice constant is unity. The employment of this form of Coulomb energy is equivalent 
to the introduction of the cut-off parameter in one-dimensional systems to avoid the 
divergence and corresponds to the screening around the on-site. [4] 
The excitonic states are obtained as the lowest-energy bound eigenstates of this Hamiltonian. 
The exciton binding energy is calculated as the difference of energy between the lowest energy 
states with and without the Coulomb energy. The Hamiltonian is numerically diagonalized by 
the Lanchos method for Kagome lattice of the finite size as large as $18 \times 18$. 
Since we are interested in qualitative features of excitons, in the calculation, the 
parameters are chosen adequately as $U_{0}$=0.1, $U_{1}$=0.75$U_{0}$, and $t^{a}_{mn}=
t^{b}_{mn}=-1.0.$ [4]

\section{Caluclated results and Disscussions}

\subsection{Binding energy of exciton}
 First, we consider the exciton binding energy on the Kagome lattice. In order to characterize 
the Kagome system, we compare binding energies among various lattices. Two-dimensional 
Kagome, square, and triangle lattices and one-dimensional one are shown in Figs.1(a) to 1(c) 
as solid lines, respectively. It should be noticed here that, by introducing transfer energies, 
$t^{\prime}$, along broken lines in Figs.1(b) and 1(c) and gradually changing their valves from 
$t^{\prime}=0$ to $t^{\prime}=-1$, one can obtain Kagome and triangle lattices from one-dimensional 
and Kagome ones, respectively. This treatment enables us to study the effect of the continuous dimension 
reduction of a lattice on the exciton binding energy from two to one by way of a Kagome lattice. 
In the followings, we also consider these lattice systems between triangle and one-dimensional lattices 
with varying the transfer energies, $t^{\prime}$ .

\begin{figure} 
\includegraphics[width=7cm,clip]{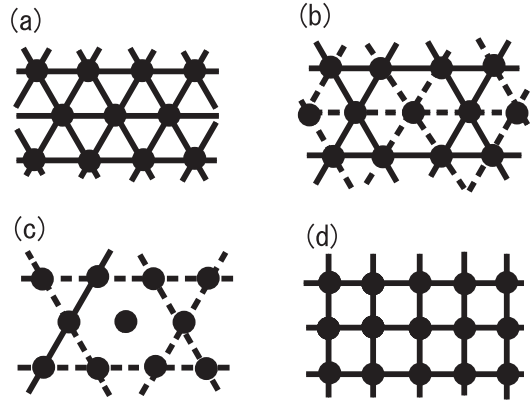} 
\caption{The lattice models adopted in this work;
(a) triangle, (b) Kagome, (c) one-dimensional, and (d) square lattices are shown by solid lines. 
Kagome and one-dimensional lattices are obtained from triangle and Kagome ones, respectively, 
by removing transfer energies shown by broken lines in (b) and (c). } 
\label{fig:geom} 
\end{figure}

 Figure 2 shows the calculated binding energies of exciton for various lattices. 
It is well known that the exciton binding energy becomes large as the dimensionality of the system decreases. 
As expected from this result, the binding energies in two-dimensional triangle and square lattices are smaller 
than one-dimensional lattice. However, in spite that the bond connection decreases in a one-dimensional 
lattice compared to a Kagome one, the exciton binding energy in the Kagome lattice is much larger than 
the one-dimensional one. To clarify the excitonic feature furthermore, the calculated exciton wave functions 
are shown in Fig.3 for Kagome, one-dimensional, and triangle lattices. This figure represents the spatial 
distribution of a hole as a function of the distance, $r$, between a hole and an electron 
when the electron is fixed on the $r$=0 site, where the distance, $r$, is measured in 
units of lattice constant and taken along one direction. As seen in Fig.3 the excitonic
radius is extremely small, around two lattice sites, in the Kagome lattice, being much 
smaller than the other lattices discussed in this paper. Apparently, these results reflect the 
localized nature of flat-band states, i.e. the conduction-band electron and valence-band hole 
states in the Kagome lattice.

\begin{figure} 
\includegraphics[width=7cm,clip]{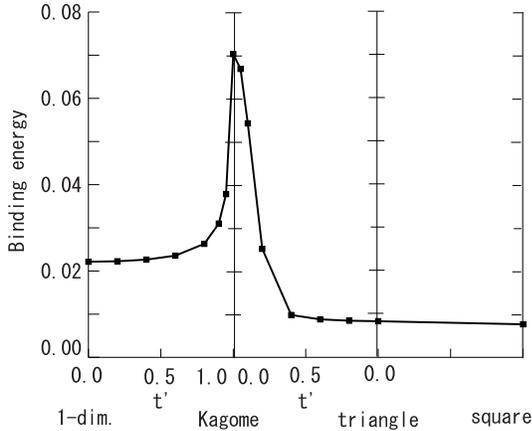} 
\caption{Calculated binding energies of exciton for various lattices. $t^{\prime}$corresponds to transfer energies 
along broken lines shown in Figs.1(b) and 1(c), and continuously varies, yielding the modification from 
the one-dimensional and Kagome lattices to Kagome and triangle lattices, respectively. 
Calculations are performed for $18 \times 18$ unit.} 
\label{fig:geom} 
\end{figure}

\begin{figure} 
\includegraphics[width=7cm,clip]{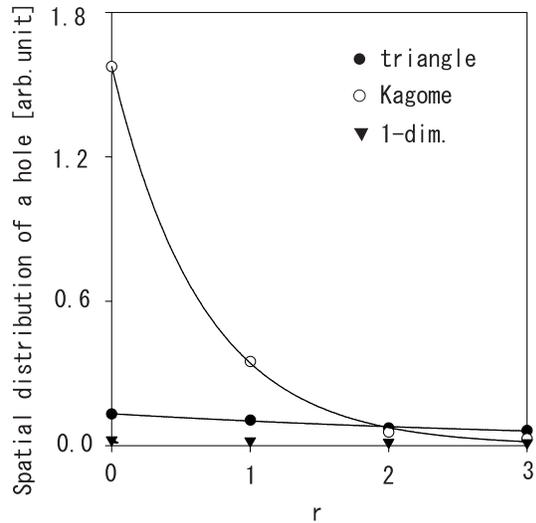} 
\caption{Calculated exciton wave functions for various lattices as a function of the distance 
between a hole and an electron, in the unit of lattice constant. 
Calculations are performed for $3 \times 3$ unit.} 
\label{fig:geom} 
\end{figure}

 Next, we consider the variation of exciton binding energy when the magnetic field is applied 
perpendicular to the surface. In the tight-binding model, the magnetic field effect is introduced 
into the Hamiltonian by multiplying the transfer energy $t_{mn}$ by the phase factor, 
$\exp[i\frac{2\pi e}{hc} \int_n^m \bm{A} \cdot d\bm{r}]$ ,where $\bm{A}$ is the vector 
potential. In the case of no Coulomb interaction, with increasing the magnetic field, 
the flat band becomes dispersive and its position leaves from the conduction-band bottom to 
the conduction-band top, which also applies to the valence band. 
Figure 4 shows the calculated exciton binding energy as a function of magnetic field, 
where the magnetic field corresponds to the flux in the unit cell of Kagome lattice and is 
measured in the unit of flux quantum. In insert, we also display the schematic band 
dispersion of the conduction band in the cases of zero and one flux quantum. It is seen that 
the exciton binding energy is the largest in the case of no magnetic field and suddenly 
decreases with applying the magnetic field. This indicates that one can largely control the 
binding energy of exction by using a Kagome lattice, and the flatness of the lowest-conduction 
and highest-valence bands is essential to the large exciton binding energy. Moreover, we note 
that the binding energy with magnetic field shows complicated variation and changes 
between those of one and two-dimensional lattices, the reason of which has not been clarified yet. 

\begin{figure} 
\includegraphics[width=7cm,clip]{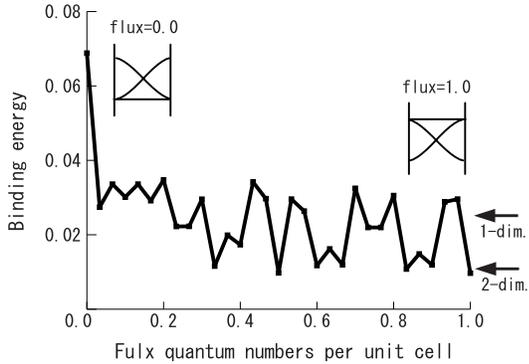} 
\caption{Calculated exciton binding energy as a function of magnetic field perpendicular to 
the Kagome lattice plane. Calculations are performed for $15 \times 15$ unit. 
Binding energies of two-dimensional (2-dim) triangle and one-dimensional (1-dim) lattices are 
denoted by arrows.} 
\label{fig:geom} 
\end{figure}

\subsection{Stability of exciton}
 Then we consider the stability of excitonic state. Since the radius of flat-band exciton is 
extremely small as shown above, the high exciton density is expected when the system is 
highly excited. Thus, we consider the stability of exciton by calculating the binding energies 
of charged exciton and biexciton, which are made of two electrons and one hole, and two 
electrons and two holes, respectively. The binding energies of these exciton complexes are 
calculated in a similar way to that of exciton, where the similar Coulomb repulsion interaction 
is introduced between electrons or holes. Moreover, we treat electrons and holes as spinless 
fermions, thus the exciton complexes corresponding to spin-singlet states. The calculated 
binding energies of the exciton, charged exciton and biexceion are shown in Fig.5. It is seen 
that the biexciton is unstable, while the charged exciton becomes stable. This might be related 
to the localized nature of carriers in a Kagome lattice. However, the detailed analysis is now 
processing and will be published elsewhere.

\begin{figure} 
\includegraphics[width=7cm,clip]{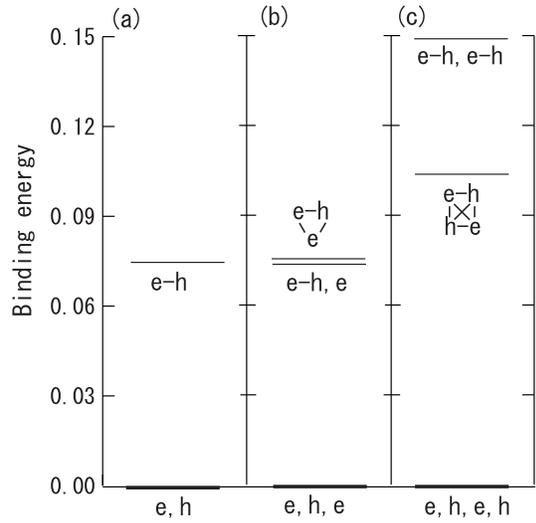} 
\caption{Calculated binding energies of 2, 3, and 4 particle states in the Kagome lattice. 
(a) States with one electron and one hole, (b) states with two electrons and one hole, 
and (c) states with two electrons and two holes. In (b), for examples, e-h,e schematically 
represents that one electron and one hole produce a bound state, while the other electron is not bounded. 
Calculations are performed for $3 \times 3$ unit.} 
\label{fig:geom} 
\end{figure}

\section{Conclusion}
 We have investigated the excitonic properties in Kagome lattice system by using the exact 
diagonalization of a tight binding model. The following features are clarified. (1) 
The exciton binding energy in the Kagome lattice is much larger than those in another 
two-dimensional systems, such as triangle and square lattices, and even the one-dimensional 
one. Furthermore the exciton radius is extremely small in the Kagome lattice, compared with 
the other lattices discussed in this paper. (2) The exciton binding energy is the largest in the 
case of no magnetic field, and suddenly decreases with applying the magnetic field. (3) 
When we treat electrons and holes as spinless fermions, the biexciton is unstable, 
while the charged exciton becomes stable.

\section{Acknowledgement}
The authors would like to thank Dr. K. Ishida for many helpful discussions. 

\section{Reference}

[1] C. Albrecht, J .H. Smet, K. von Klizing, D. Weiss, V. Umansky and H. Schweizer, 
Phys. Rev. Lett. \textbf{86}, (2001) 147.

\begin{flushleft}
[2] A. Mielke and H. Tasaki, Commun. Math. Phys. \textbf{158}, (1993) 341.
\end{flushleft}

\begin{flushleft}
[3] K. Shiraishi, H. Tamura and H. Takayanagi, Appl. Phys. Lett. \textbf{78}, (2001) 3702.
\end{flushleft}

\begin{flushleft}
[4] K. Ishida, Phys. Rev. B. \textbf{49}, (1994) 5541.
\end{flushleft}

\end{document}